\documentclass[fleqn,usenatbib]{mnras}

\usepackage{newtxtext,newtxmath}
\usepackage{xcolor}

\usepackage[T1]{fontenc}

\DeclareRobustCommand{\VAN}[3]{#2}
\let\VANthebibliography\thebibliography
\def\thebibliography{\DeclareRobustCommand{\VAN}[3]{##3}\VANthebibliography}

\usepackage{graphicx}	
\usepackage{amsmath}	
\usepackage{ulem}

\title[MG fits to 6dFGS and SDSS velocity data]{Constraining modified gravity scenarios with the 6dFGS and SDSS galaxy peculiar velocity datasets}

\author[Lyall et al.]{
Stuart Lyall$^{1}$\thanks{E-mail: slyall@swin.edu.au},
Chris Blake$^{1}$, and
Ryan J. Turner$^{1}$
\\
$^{1}$ Centre for Astrophysics \& Supercomputing, Swinburne
  University of Technology, P.O.\ Box 218, Hawthorn, VIC 3122,
  Australia
}

\date{Accepted XXX. Received YYY; in original form ZZZ}

\pubyear{2023}

\begin{document}
\label{firstpage}
\pagerange{\pageref{firstpage}--\pageref{lastpage}}
\maketitle

\begin{abstract}
The detailed nature of dark energy remains a mystery, leaving the possibility that its effects might be explained by changes to the laws of gravity on large scales.  The peculiar velocities of galaxies directly trace the strength of gravity on cosmic scales and provide a means to further constrain such models. We generate constraints on different scenarios of gravitational physics by measuring peculiar velocity and galaxy clustering two-point correlations, using redshifts and distances from the 6-degree Field Galaxy Survey and the Sloan Digital Sky Survey Peculiar Velocity samples, and fitting them against models characteristic of different cosmologies. Our best-fitting results are all found to be in statistical agreement with General Relativity, in which context we measure the low-redshift growth of structure to be $f\sigma_8 = 0.329^{+0.081}_{-0.083}$, consistent with the prediction of the standard $\Lambda$CDM model.  We also fit the modified gravity scenarios of Dvali-Gabadadze-Porrati (nDGP) and a Hu-Sawicki model of $f(R)$ gravity, finding the $2\sigma$ limit of their characteristic parameters to be $r_cH_0/c>6.987$ and $-\log_{10}(|f_{R0}|)>4.703$, respectively. These constraints are comparable to other literature values, though it should be noted that they are significantly affected by the prior adopted for their characteristic parameters. When applied to much larger upcoming peculiar velocity surveys such as DESI, this method will place rapidly-improving constraints on modified gravity models of cosmic expansion and growth.
\end{abstract}




\section{Introduction}

The Universe has long been observed to be expanding over time. Current cosmological measurements show further that the rate of expansion is increasing \citep[for a review, see][]{2013PhR...530...87W}. This observation can not be explained by Einstein's General Relativity (GR) applied to a homogeneous and isotropic universe made up of currently-understood matter-energy components alone.

The current standard model for universal expansion, the $\Lambda$CDM cosmological model, introduces a cosmological constant component $\Lambda$.  This model fits observations of the cosmic expansion history quite well \citep[e.g.,][]{2018ApJ...859..101S, 2020A&A...641A...6P, 2021PhRvD.103h3533A} although there remain tensions in determinations of the Hubble constant $H_0$ by different methods \citep[e.g.,][]{2021ApJ...919...16F, 2021CQGra..38o3001D, 2022ApJ...934L...7R}. The cosmological constant $\Lambda$ represents a special form of `dark energy' with an equation of state of $w=-1$, producing a repulsive effect on cosmic scales.

Although the phenomenological model for dark energy incorporated in the $\Lambda$CDM model successfully fits many cosmological observations, it currently offers no theoretical insight into the underlying nature of dark energy. In alternative attempts to explain cosmic expansion history, a number of modified gravity models have been proposed \citep[for reviews, see][]{2012PhR...513....1C, 2019LRR....22....1I, 2019ARA&A..57..335F}.  Two of the most prominent models under consideration, which we focus on in the current study, are the Dvali-Gabadadze-Porrati (DGP) and $f(R)$ models.

The DGP models introduce interactions with higher-dimensional manifolds \citep{2000PhLB..485..208D}. For the normal branch models \citep{2003JCAP...11..014S,2004PhRvD..70j1501L}, which have yet to be fully disproven, the strength of this interaction can be represented by the cross-over length scale parameter $r_c$. The Cosmic Microwave Background (CMB) measurements provide the constraint $r_c > 3.5 c/H_0$ to 95\% confidence \citep{2009PhRvD..80f3536L}. Other measurements of the cross-over scale have been performed using large-scale structure redshift-space distortion and distance-scale datasets, finding $r_c \gtrsim 1 c/H_0$ \citep[e.g.,][]{2013MNRAS.436...89R,2016PhRvD..94h4022B}.

The $f(R)$ models seek to investigate a more general space of gravitational interaction terms that could evidence other potential quantum models of gravity \citep[][]{2004PhRvD..70d3528C,2007PhRvD..76f4004H,2010RvMP...82..451S}. The Hu-Sawicki models \citep{2007PhRvD..76f4004H}, which we focus on in this paper, can be parameterised by the interaction strength of its divergent action term ($f_{R0}$). Recent constraints from cosmological analyses approximately yield $\log_{10}|f_{R0}| \lesssim -5$ \citep[e.g.][]{2012PhRvD..85l4038L,2015PhRvD..92d4009C,2016PhRvL.117e1101L}. Solar system and astrophysical tests on smaller scales are more constraining, producing $\log_{10}|f_{R0}| \lesssim -6$ \citep[e.g.][]{2007PhRvD..76f4004H,2013ApJ...779...39J,2014arXiv1409.3708S,2020PhRvD.102j4060D}.


A promising new probe of these scenarios is galaxy peculiar velocities obtained using redshift-independent distance measurements \citep{1995PhR...261..271S}. At cosmic distance scales the average residual motion of galaxies through space can be modelled linearly to high accuracy. This effectively makes the peculiar velocity of a galaxy a statistical tracer of the gravitational field strength it experiences. Modified gravity models, like those mentioned above, seek to reconstruct large-scale cosmic acceleration. This means that peculiar velocity probes are sensitive to changes in the strength of gravity on the scales that we expect modified gravity models to measurably diverge from General Relativity \citep{2004MNRAS.347..255B, 2014MNRAS.445.4267K, 2017MNRAS.464.2517H, 2020MNRAS.497.1275S, 2023MNRAS.518.5929L}.

Through linear theory, galaxy peculiar velocities can be linked to the growth rate of structure ($f$), the rate of growth of matter density perturbations, allowing observable peculiar velocity correlations to be parameterised by this variable.  The effect is commonly measured by the degenerate parameter pair $f\sigma_8$, where $\sigma_8$ describes the normalisation of the matter power spectrum \citep{2023A&ARv..31....2H}.  The growth rate of structure can also be calculated from linear perturbation theory for generalised gravity models as a function of scale \citep[e.g.,][]{2016PhRvD..94h4022B, 2019JCAP...09..066M}, allowing theory to be compared to observable results for a range of scenarios.

The largest current single observational datasets that can be used for galaxy peculiar velocity studies are the 6-degree Field Galaxy Survey (6dFGS) peculiar velocity sample \citep{2014MNRAS.445.2677S} and the Sloan Digital Sky Survey (SDSS) peculiar velocity sample \citep{2022MNRAS.515..953H}, which both use the Fundamental Plane method for determining redshift-independent distances.  6dFGS measured approximately one hundred thousand spectroscopic redshifts including approximately $9{,}000$ distances, whilst SDSS has measured several hundred thousand spectroscopic redshifts and approximately $34{,}000$ distances.  These datasets are located in non-overlapping portions of the sky, such that their results can be independently combined.  We note that larger compilations of peculiar velocity datasets have been presented in the {\it Cosmicflows} catalogues \citep{2016AJ....152...50T, 2020ApJ...902..145K, 2023ApJ...944...94T}.  These are heterogeneous samples, for which it is harder to construct matched mock catalogues for testing our analysis pipelines.

The growth of structure within peculiar velocity datasets has been analysed by numerous previous studies.  Focusing first on studies of the 6dFGS PV sample used in our paper, \cite{2014MNRAS.444.3926J} fit the correlation of the peculiar velocity field of the 6dFGS and a sample of supernovae, finding $f\sigma_8=0.418\pm0.065$; \cite{2020MNRAS.494.3275A} performed a joint maximum-likelihood study of the overdensity and velocity fields of 6dFGS, determining $f\sigma_8 = 0.384 \pm 0.052$; and \cite{2023MNRAS.518.2436T} used density and velocity correlation-function fitting on the 6dFGS dataset and found a value of $f \sigma_8 = 0.358 \pm 0.075$.  These results are all statistically consistent with a standard GR+$\Lambda$CDM prediction for the sample, $f\sigma_8 \approx 0.42$ (where the exact prediction depends on the $\Omega_m$ value).  Regarding studies of the SDSS sample, \cite{2023MNRAS.518.1840L} performed a maximum-likelihood analysis of the density and velocity fields, determining $f\sigma_8=0.405^{+0.076}_{-0.071}$.

Related growth rate analyses of other samples, including the {\it Cosmicflows} catalogues and compilations of Type Ia supernovae, have been presented by, for example, \cite{2005ApJ...635...11P, 2011MNRAS.413.2906D, 2015MNRAS.450..317C, 2017MNRAS.471.3135H, 2017JCAP...05..015H, 2017MNRAS.470..445N, 2019MNRAS.486..440D, 2019MNRAS.487.5235Q, 2020MNRAS.498.2703B, 2020MNRAS.497.1275S, 2023A&A...670L..15C}.  These studies present a range of methodologies, sometimes including additional modelling of the velocity field, which is then compared with the observed peculiar velocities.  The results of these studies generally agree with the standard GR+$\Lambda$CDM growth rate prediction, with a typical fractional error of $\sim 20\%$, although \cite{2020MNRAS.497.1275S} recover a lower growth rate value.

In the current study we extend these analyses in two ways.  First we perform the first cosmological analysis of the galaxy and velocity correlation functions of the SDSS PV catalogue \citep{2022MNRAS.515..953H}, allowing us to compare with the growth rate determinations of the analogous analysis of the 6dFGS PV sample by \cite{2023MNRAS.518.2436T}.  Second, we constrain modified gravity models using both the 6dFGS and SDSS samples, building on the simulation study we presented in \cite{2023MNRAS.518.5929L}.  To our knowledge, these are the first constraints on DGP and $f(R)$ models from galaxy peculiar velocities.  These measurements are expected to improve rapidly in the future, given that the size of these datasets is expected to increase by more than an order of magnitude over the next few years with the advent of new peculiar velocity datasets from the Dark Energy Spectroscopic Instrument \citep{2023MNRAS.525.1106S}, the 4-metre Multi-Object Spectroscopic Telescope (4MOST) Hemisphere Survey \citep{2023Msngr.190...46T}, the Vera Rubin Observatory \citep{2017ApJ...847..128H} and the Australian Square Kilometre Array Pathfinder WALLABY survey \citep{2023MNRAS.519.4589C}.

This paper is structured as follows.  In Sec.\ref{sec:theory} we outline the background theory, summarising the modified gravity models we consider and the predicted correlation functions.  Sec.\ref{sec:data} then describes the data being utilised and the method of analysis. The results of fitting a $\Lambda$CDM model to these data will be shown in Sec.\ref{sec:gr}. Sec.\ref{sec:modgrav} will show the fitting results in the context of the modified gravity models we consider, before we conclude in Sec.\ref{sec:conc}.

\section{Theory}
\label{sec:theory}

\subsection{Growth in MG scenarios}
\label{sec:modgrowth}

The growth rate of structure $f$, is a cosmological variable that represents the rate of gravitational development of matter overdensities as the universe evolves \citep[for a recent review, see][]{2023A&ARv..31....2H}. It is defined by:
\begin{equation}
f=\frac{d\ln(|\delta_m|)}{d\ln(a)} ,
\label{eq_growth}
\end{equation}
where $\delta_m$ is the matter overdensity, and $a$ is the cosmic scale factor.  While overdense regions of the universe will collapse, underdense regions will have more matter pulled from them. Through gravitational acceleration being proportional to gravitating mass, the rate of change of the overdensity for any region is proportional to the current overdensity in the linear approximation. It follows that in the linear approximation, the logarithmic derivative of overdensity can be treated as a global variable.

In the linear theory regime, peculiar velocities $\mathbf{v}(\mathbf{x})$ become a tracer of gravity. Correlations between peculiar velocities and matter perturbations can be used to constrain the growth rate of structure via the standard equation \citep[for a derivation, see e.g.][]{2020MNRAS.494.3275A}:
\begin{equation}
\mathbf{\nabla} \cdot \mathbf{v}(\mathbf{x}) = -aHf\delta_m ,
\label{eq_vdiv}
\end{equation}
where $H$ is the Hubble parameter. This equation explicitly shows that peculiar velocities can be used to trace large-scale structure, providing an observable probe of the large-scale effects of gravity which can help distinguish between modified gravity models.

The leading current cosmological model is the $\Lambda$CDM model. This model uses Einstein's Theory of General Relativity to describe the evolution of the entire universe. The field equations dictating the motion of matter, energy, and spacetime can be derived from the Einstein-Hilbert action:
\begin{equation}
S=\int\frac{R-2\Lambda}{16\pi G}\sqrt{|g|}dx^4 ,
\label{eq_EHaction}
\end{equation}
where $G$ is the gravitational constant, $R$ is the Ricci scalar describing the curvature of spacetime, $|g|$ is the determinant of the spacetime metric or the volume element of spacetime, and $\Lambda$ is the cosmological constant which records the density of dark energy.

The action is a convenient formalism for compactly expressing the dynamics of a system and becomes a natural starting point for modifications to the theory.  Here we focus on the DGP and $f(R)$ models, which are commonly used as representative examples of alternative theories \citep{2012PhR...513....1C, 2016ARNPS..66...95J, 2019LRR....22....1I, 2019ARA&A..57..335F}. In the DGP scenario a term is included describing an interaction with a higher-dimensional spacetime curvature:
\begin{equation}
S=\int\frac{R^{(5)}}{16\pi G^{(5)}}\sqrt{|g|^{(5)}}dx^5+\int\bigg(\frac{R}{16\pi G}+\mathcal{L}^{(m)}\bigg)\sqrt{|g|}dx^4 ,
\end{equation}
where the theory is parameterised by the crossover length,
\begin{equation}
r_c=\frac{1}{2}\frac{G^{(5)}}{G} .
\end{equation}
In the $f(R)$ scenario the action becomes,
\begin{equation}
S=\int\frac{f(R)}{16\pi G}\sqrt{|g|}dx^4 ,
\label{eq_fRaction}
\end{equation}
where we assume the Hu-Sawicki formulation \citep{2007PhRvD..76f4004H},
\begin{equation}
f(R)=R-2\Lambda-f_{R0}\frac{R_0^2}{R} ,
\end{equation}
where the theory is parameterised by the amplitude $f_{R0}$.

Reparameterising the linear matter perturbation evolution equations in terms of the growth rate of structure provides an equation which may be generalised across different modified gravity scenarios \citep[e.g.][]{2005PhRvD..72d3529L,2023MNRAS.518.5929L},
\begin{equation}
\frac{df}{da}=\frac{3G_{\rm eff} H_0^2 \Omega_{m0}}{2 a^4 H^2}-\left(\frac{1}{H}\frac{dH}{da}+\frac{2}{a}\right)f-\frac{f^2}{a} ,
\label{eq_linpertf}
\end{equation}
where $\Omega_{m0}$ and $H_0$ are the current values of the matter density parameter and Hubble parameter, respectively, and $G_{\rm eff}$ is the effective strength of gravity. This parameter provides a single function that encodes all the altered effects of using a modified gravity theory.  In GR,
\begin{equation}
G_{\rm eff}=1 .
\end{equation}
For DGP the gravity strength factor becomes \citep{2016PhRvD..94h4022B},
\begin{equation}
G_{\rm eff}=1+\frac{1}{3 \left[ 1+2\frac{r_c}{c}\left( H+\frac{1}{3}\frac{dH}{d\ln(a)} \right) \right]} ,
\end{equation}
and for $f(R)$ the factor becomes \citep{2019JCAP...09..066M},
\begin{equation}
G_{\rm eff}(k) = \left(\frac{df(R)}{dR}\right)^{-1}\left(1+\frac{1}{\frac{a^2}{k^2}\left(\frac{d^2f(R)}{dR^2}\right)^{-1}+3}\right) ,
\label{eq_GeffR}
\end{equation}
where $k$ is the wavenumber of the Fourier density mode.  As can be seen, the growth rate becomes scale-dependent in $f(R)$ models.  In this sense, the growth of structure is an important point of comparison between theory and observations.

The cumulative amplitude of a growing perturbation across time is parameterised by the growth factor $g$, which is given by inverting Eq.\ref{eq_growth},
\begin{equation}
g(k,a) = g_0 \, \exp{\left[ -\int_{a_{\rm ini}}^a \frac{f(k,a')}{a'} \, da' \right]} ,
\label{eq_growfac}
\end{equation}
Once the growth rate of structure has been determined by Eq.\ref{eq_linpertf}, it can be used to determine the growth factor. The growth factor also has a normalisation factor $g_0$, which is tied to the overall normalisation of the matter power spectrum.  In our convention this will simply take the value $g_0 = 1$ for GR models, although it must be marginalised in general, as we discuss in Sec.\ref{sec:mg}.

\subsection{Correlation function models}
\label{sec:corrmod}

Cosmological measurements indicate that the universe is homogeneous and isotropic on large scales. Hence, we expect the average value of an observable to be independent of large-scale position and orientation (neglecting redshift-space distortions for the moment, which we will discuss below). This implies that relationships between an object and surrounding quantities can be sufficiently modelled as a 2-point correlation function that only depends on separation.

As peculiar velocities can be related to the growth of structure, a correlation that we expect to be sensitive to gravity is the velocity-velocity auto-correlation function,
\begin{equation}
    \xi_{vv}^{\alpha\beta}(r) = \left \langle v^\alpha(\mathbf{x}) \, v^\beta(\mathbf{x}+\mathbf{r}) \right \rangle ,
\end{equation}
where $\mathbf{v}$ is the peculiar velocity, $\mathbf{x}$ is the galaxy position and $\mathbf{r}$ is the separation vector. $\xi_{vv}^{\alpha\beta}$ encodes information about the correlation strength and scale of the bulk-flow movement of galaxies along relative directions $\alpha$ and $\beta$.  In practical observational samples, we can only measure the radial velocity projected along the line-of-sight, which alters the accessible correlation statistics as we discuss below.

Further information detailing structure growth can be gained from the galaxy-velocity cross-correlation function,
\begin{equation}
    \xi_{gv}(r)=\left\langle\delta_m(\mathbf{x}) \, (\mathbf{v}(\mathbf{x}+\mathbf{r})\cdot(-\mathbf{\hat{r}}))\right\rangle .
\end{equation}
$\xi_{gv}$ communicates the infall rate of galaxies towards overdense regions.  Finally, our study will make use of the galaxy-galaxy auto-correlation function,
\begin{equation}
    \xi_{gg}(r)=\left\langle\delta_m(\mathbf{x}) \, \delta_m(\mathbf{x}+\mathbf{r})\right\rangle ,
\end{equation}
which captures the scale-dependent amplitude of the clustering of galaxies. This correlation is not directly sensitive to the rate of structure growth in real space, but it does help constrain free parameters in the model, including galaxy bias.

The assumption of isotropy holds in real space distributions, but in practice redshift-space distortion (RSD) creates a dependence of the physics on the angle between the separation vector and line of sight.  RSD effects are imprinted because the positions of galaxies are inferred from their measured redshifts, which are biased by their peculiar velocities \citep{1987MNRAS.227....1K}.  This effect, whilst complicating our analysis, encodes additional information about the growth rate of structure which can be used to test our models.

In redshift space, the correlation functions between objects depend on the angle $\theta$ of the separation vector of two objects with respect to the line of sight, which we parameterise as $\mu = \cos{\theta}$.  These dependences on $\mu$ are expressed as multipole components using a Legendre polynomial expansion. This deconstruction is convenient as linear theory predicts that all useful information about the model can be found in the first few multipole modes \citep[e.g.][]{2020MNRAS.494.3275A, 2023MNRAS.518.1840L}. This means that we can capture all usable information from our three real-space correlation functions in five redshift-space correlation functions \citep{2023MNRAS.518.2436T}.

These correlation functions can be theoretically modeled in the linear regime using a known cosmological model, matter power spectrum and gravity law. First, we utilise the monopole and quadrupole of the galaxy auto-correlation:
\begin{equation}
    \xi_{gg}^{0}(r)=\frac{1}{2\pi^2}\int k^2 \, j_0(kr) \, M^0_{gg}(k) \, dk ,
\label{eq:gg0corr}
\end{equation}
\begin{equation}
    \xi^2_{gg}(r)=\frac{1}{2\pi^2}\int k^2 \, j_2(kr) \, M^2_{gg}(k) \, dk ,
\label{eq:gg2corr}
\end{equation}
(where we will shortly define the terms appearing).  Then, the dipole of the galaxy-velocity cross-correlation:
\begin{equation}
    \xi^1_{gv}(r)=-\frac{aH}{2\pi^2}\int k \, j_1(kr)\left(M^0_{gv}(k)+\frac{2}{5}M^2_{gv}(k)\right)dk .
\label{eq:gv1corr}
\end{equation}
Finally, we decompose the velocity auto-correlation into two functions equivalent to the monopole and quadrupole of the radial velocity correlation \citep{1989ApJ...344....1G}:
\begin{equation}
    \psi_\parallel(r)=\frac{H^2a^2}{2\pi^2}\int M^0_{vv}(k)\left(j_0(kr)-\frac{2j_1(kr)}{kr}\right)dk ,
\label{eq:vvparcorr}
\end{equation}
\begin{equation}
    \psi_\perp(r)=\frac{H^2a^2}{2\pi^2}\int M^0_{vv}(k)\left(\frac{j_1(kr)}{kr}\right)dk .
\label{eq:vvperpcorr}
\end{equation}

In the above equations, $j_n$ is the spherical Bessel function of rank $n$, and $M^\ell_{xy}$ is the modified power spectrum multipole component $\ell$ for the correlation $\xi_{xy}$, given by:
\begin{equation}
    M^\ell_{gg}(k)=\frac{2l+1}{2}\int_{-1}^1(b+f\mu^2)^2 \, D_g^2(k,\mu) \, P(k) \, L_\ell(\mu) \, d\mu ,
\label{eq:ggpower}
\end{equation}
\begin{equation}
    M^\ell_{gv}(k)=\frac{2l+1}{2}\int_{-1}^1(b+f\mu^2) \, f \, D_g(k,\mu) \, D_v(k) \, P(k) \, L_\ell(\mu) \, d\mu ,
\label{eq:gvpower}
\end{equation}
\begin{equation}
    M^\ell_{vv}(k)=\frac{2l+1}{2}\int_{-1}^1 \, f^2 \, D_v^2(k) \, P(k) \, L_\ell(\mu) \, d\mu .
\label{eq:vvpower}
\end{equation}
Here, $P(k)$ is the matter power spectrum, $L_\ell(\mu)$ is the Legendre polynomial of mode $\ell$, $b$ is the linear galaxy bias, $f$ is the growth rate of structure (which can depend on scale in some models), and $D_x$ are damping terms used to model non-linear effects. We introduce these variables now.

The linear galaxy bias $b$ represents the first-order relationship between the observable galaxy overdensity distribution $\delta_g$ and the total matter overdensity distribution $\delta_m$,
\begin{equation}
    \delta_g = b \, \delta_m + O(\delta_m^2) .
\end{equation}
The damping term $D_g$ represents the non-linear effects of galaxy velocities in RSD correlation functions involving galaxies, which we parameterise as \citep{1998MNRAS.296...10H},
\begin{equation}
    D_g(k,\mu)=\frac{1}{\sqrt{1+(k\mu\sigma_v/H_0)^2}} ,
\end{equation}
where $\sigma_v$ encapsulates this velocity dispersion, which is a free parameter in our model.  The damping term $D_v$ represents the non-linear effects in RSD correlation functions involving velocities \citep{2014MNRAS.445.4267K},
\begin{equation}
    D_v(k)=\frac{\sin(k\sigma_u)}{k\sigma_u} ,
\end{equation}
where $\sigma_u$ is typically determined via simulations as discussed below.

Finally, the factor $P(k)$ in Eq.\ref{eq:ggpower} to Eq.\ref{eq:vvpower} represents the matter power spectrum at the observation redshift, which we allow to be distorted by modified gravity effects according to an approximate formulation \citep[following][]{2023MNRAS.518.5929L}.  We start by calculating the GR matter power spectrum $P_{GR}$ with the python {\tt camb} package \citep{2000ApJ...538..473L}. In the linear regime, the evolution of the power spectrum is described by the square of the growth factor (Eq. \ref{eq_growfac}). Hence, the relative power spectrum for a modified gravity model is given by the ratio,
\begin{equation}
P(k,a) = P_{GR}(k,a) \, \frac{g^2(k,a)}{g^2_{GR}(a)} .
\label{eq_powerMG}
\end{equation}
With the equations given in this section, we can model the five observable correlation functions (Eq.\ref{eq:gg0corr} - \ref{eq:vvperpcorr}) for a given gravity model. The four free variables for these correlation functions are: the characteristic model specific parameter that controls the deviation in growth of structure, $b$: the linear galaxy bias, $\sigma_v$: the velocity dispersion, and $g_0$: the normalisation of the power spectrum.  We vary these parameters in our analysis to explore their likelihood relative to the data.

Since we can only measure the line-of-sight components of velocities, the theoretical velocity correlation functions $\psi_\parallel$ and $\psi_{\perp}$ can not be directly measured. To capture the information in these correlations observationally, we use the velocity auto-correlation estimators $\psi_1$ and $\psi_2$, which are defined in Sec.\ref{sec:corrmeas} below. These observable correlations are theoretically predicted by \citep{1989ApJ...344....1G}:
\begin{equation}
\begin{split}
    \psi_1(r) &= \mathcal{A}(r) \, \psi_\parallel(r) + \left[ 1-\mathcal{A}(r) \right] \, \psi_\perp(r) , \\
    \psi_2(r) &= \mathcal{B}(r) \, \psi_\parallel(r) + \left[ 1-\mathcal{B}(r) \right] \, \psi_\perp(r) ,
\end{split}
\end{equation}
where $\mathcal{A}$ and $\mathcal{B}$ are dataset-dependent functions given as \citep{2023MNRAS.518.2436T}:
\begin{equation}
\begin{split}
    \mathcal{A}(r) = \frac{\sum_{a,b} w_a w_b \, \cos\theta_a \, \cos\theta_b \, \cos\theta_{ab}}{\sum_{a,b} w_a w_b \, \cos^2\theta_{ab}} , \\
    \mathcal{B}(r) = \frac{\sum_{a,b} w_a w_b \cos^2\theta_a \, \cos^2\theta_b}{\sum_{a,b} w_a w_b \, \cos\theta_a \, \cos\theta_b \, \cos\theta_{ab}} ,
\end{split}
\end{equation}
where the sum is over all pairs of data points $a$ and $b$ in each bin, the galaxy weights $w_a$ and $w_b$ are defined in Sec.\ref{sec:corrmeas}, and the angles with respect to the line-of-sight are given by $\cos\theta_a = \mathbf{\hat{x}}_a \cdot \mathbf{\hat{r}}$ and $\cos\theta_{ab} = \mathbf{\hat{x}}_a \cdot \mathbf{\hat{x}}_b$, where $\mathbf{\hat{x}}_a$ is the normalised position vector of galaxy $a$. These angles are depicted by Fig.1 in \cite{2023MNRAS.518.2436T}.

\section{Data}
\label{sec:data}

We fit these models to datasets from two peculiar velocity (PV) surveys: the 6-degree Field Galaxy Survey PV sample \citep{2014MNRAS.445.2677S} and the Sloan Digital Sky Survey PV sample \citep{2022MNRAS.515..953H}. These two datasets constitute the current largest homogeneous set of velocity measurements.  In the following sections, we briefly summarise these datasets and the corresponding mock catalogues we use in each case to validate our analysis.

\subsection{6dFGS sample}

The 6-degree Field Galaxy Survey (6dFGS) provides a catalogue of galaxy redshifts and peculiar velocities across a roughly $17{,}000 \, {\rm deg}^2$ region of the southern sky \citep{2009MNRAS.399..683J}. The original survey was conducted between 2001 and 2006 by the UK Schmidt Telescope, a 1.24 metre telescope situated at the Siding Spring Observatory.

We draw our density-field sample from the 6dFGS galaxy redshift catalogue originally used for baryon acoustic oscillation analysis by \cite{2011MNRAS.416.3017B}.  In this study, the original sample of $125{,}071$ spectroscopic redshifts was reduced by an apparent magnitude selection and an additional redshift cut $z < 0.1$ was applied, producing a final sample size of $70{,}467$ redshifts.  Fundamental Plane distances were measured for $11{,}287$ 6dFGS galaxies by \cite{2012MNRAS.427..245M}, which was further refined to a final velocity sample of $8{,}885$ by \cite{2014MNRAS.445.2677S}.  These same 6dFGS redshift and velocity samples were previously studied by \cite{2020MNRAS.494.3275A} and \cite{2023MNRAS.518.2436T}.

To validate our analysis and determine the data covariance as described below, we also used the 600 mock 6dFGS catalogues generated by \cite{2018MNRAS.481.2371C} with a COmoving Lagrangian Acceleration (COLA) simulation method \citep{2013JCAP...06..036T, 2015A&C....12..109H, 2016MNRAS.459.2118K}. This code seeks to drastically increase efficiency by analytically solving linear and second order dynamics whilst only calculating the nonlinear residual displacement with N-body interactions, directly trading run time for small scale resolution. Each simulation has a box-length of $1.2 \, h^{-1}$ Gpc and contains $(1728)^3$ particles with a mass resolution of $2.8 \times 10^{10} \, h^{-1} M_{\sun}$.  Dark matter halos were populated with galaxies using a halo occupation distribution model fit to the 6dFGS galaxy number density and projected correlation function.  Mock velocity sub-samples were created by selecting the most massive halos at a given redshift, representing early-type galaxies \citep{2020MNRAS.494.3275A}.

In order to match the configuration of the 6dFGS mocks we used the following set of fiducial cosmological parameters when generating the model power spectrum for 6dFGS analysis: $h = 0.68$, $\Omega_m = 0.3$, $\Omega_b = 0.0478$, $\sigma_8 = 0.82$, $n_s = 0.96$.  We also assumed an effective redshift $z=0$ and a non-linear velocity damping parameter $\sigma_u = 13 \, h^{-1}$ Mpc, following \cite{2014MNRAS.445.4267K}.  \cite{2020MNRAS.494.3275A} established that these choices do not significantly affect the cosmological conclusions, given the statistical precision of current data.

\subsection{SDSS PV sample}

The Sloan Digital Sky Survey (SDSS) is a extensive, ongoing wide-field imaging and spectroscopic survey \citep{2000AJ....120.1579Y}, using the $2.5$ metre telescope at Apache Point Observatory. The SDSS 14th Data Release (DR14) \citep{2018ApJS..235...42A} was used by \cite{2020MNRAS.497.1275S} to construct Fundamental Plane measurements, from which $34{,}059$ peculiar velocities were extracted for analysis by \cite{2022MNRAS.515..953H}, forming a sample that reaches up to $z = 0.1$.  \cite{2022MNRAS.515..953H} also presented 2048 mock simulations constructed to reproduce the clustering and selection function of the SDSS sample.  When analysing this sample, we use fiducial cosmological parameters matching these studies -- $h = 0.6751$, $\Omega_m = 0.3121$, $\Omega_b = 0.0488$, $\sigma_8 = 0.815$, $n_s = 0.9653$ -- and an effective redshift $z=0.073$.  Following the analysis of \cite{2023MNRAS.518.1840L} we adopt a fixed value $\sigma_u = 21 \, h^{-1}$ Mpc, calibrated by these simulations.

\subsection{Correlation function measurements}
\label{sec:corrmeas}

We measured the galaxy and velocity auto- and cross-correlation functions of these datasets using the same methods as described in \cite{2023MNRAS.518.2436T}.  We briefly summarise the estimators here, and refer to \cite{2023MNRAS.518.2436T} for full details.

For the peculiar velocity datasets, the radial velocity measurements are derived from independent distances determined using the Fundamental Plane technique. This method inherently produces a log-normal probability distribution for the distances to galaxies, which introduces biases in the analysis of cosmology \citep{2014MNRAS.445.2677S}.  These biases can be avoided if we instead formulate the peculiar velocity measurements in terms of the logarithmic distance ratio $\eta=\log_{10}(D(z_{obs})/D(z_H))$, where $D(z_{obs})$ is the comoving distance corresponding to the observed redshift and $D(z_H)$ is the real comoving distance. The error in this variable is well-described by a Gaussian function. Expanding $\eta$ to first order in peculiar velocities leads to \citep{2014MNRAS.444.3926J, 2020MNRAS.494.3275A}:
\begin{equation}
    \eta = \alpha(z) \, v = \frac{1}{\ln(10)}\frac{1+z_{obs}}{D(z_{obs}) \, H(z_{obs})} \, v ,
\label{eq_eta}
\end{equation}
defining the normalising factor $\alpha(z)$ which must be applied when converting correlation functions in the $\eta$ variable to the velocity correlation functions.

Our correlation function estimators are constructed from pair-counting algorithms.  We also make use of random catalogues, which are constructed to have the same survey distribution and properties as the data, but containing no clustering. $DD$, $RR$ and $DR$ refer to weighted pair counts between data points only, between random points only, and between pairs of data and random points.  We bin the correlation function measurements by the pair separation distance, $r$, and the angle with respect to the line of sight, $\mu = \cos\theta$. We use 25 separation bins of width $6 \, h^{-1}$ Mpc with a range from $0$ to $150 \, h^{-1}$ Mpc, and 20 angular bins of width $\Delta\mu = 0.1$ with a range from $-1$ to $+1$.

For the galaxy-galaxy auto-correlation function, the pair-count estimator in a bin of separation and angle has the form \citep{1993ApJ...412...64L},
\begin{equation}
    \xi_{gg}(r_i,\mu_j) = \left( \frac{N^R_g}{N^D_g} \right)^2 \frac{D_gD_g}{R_gR_g} - 2 \frac{N^R_g}{N^D_g} \frac{D_gR_g}{R_gR_g} + 1 ,
\label{eq_xigg}
\end{equation}
where the weighted pair counts are defined as,
\begin{equation}
    A_gB_g(r_i, \mu_j) = \sum_{a,b}^{A,B\in(r_i, \mu_j)} w^g_a \, w^g_b ,
\end{equation}
where the sum is taken across pairs of galaxies between data sets $A$ and $B$, where the separation and angle of the pair to the line-of-sight are within the bins $r_i$ and $\mu_j$.  The normalisation factors are the total weighted number of points in each data set,
\begin{equation}
    N^A_x=\sum_a w^x_a .
\end{equation}
The random-random pair count $R_gR_g$, appearing in the denominator of Eq.\ref{eq_xigg}, weights the other pair counts by volume and gives them physical meaning as a density. The data-random pair count $D_gR_g$ corrects for systematic survey boundary effects on the data-data pair count, reducing the variance of the correlation function \citep{1993ApJ...412...64L}.

We assign galaxies optimal weights using FKP weighting \citep{1994ApJ...426...23F}:
\begin{equation}
    w^g_a = \frac{1}{n^g_a \, P_g + 1} ,
\end{equation}
where $P_g = 10^4 \, h^{-3}\textrm{Mpc}^3$ is the characteristic galaxy power spectrum amplitude and $n^g_a$ is the local galaxy number density at galaxy $a$.  To correct the correlation function for the normalisation constraint implied by the fixed number of total pairs we add onto the estimated $\xi_{gg}$ an integral constraint given by \citep{1980lssu.book.....P},
\begin{equation}
    I.C.=\frac{\sum_{i,j} \xi_{gg}(r_i,\mu_j) \, R_gR_g(r_i,\mu_j)}{\sum_{i,j} R_gR_g(r_i,\mu_j)} ,
\end{equation}
where the sum is taken over all separation and angular bins.  We then calculate the multipole components of the measured correlation function as,
\begin{equation}
    \xi^\ell(r) = \frac{2l+1}{2} \int_{-1}^1 d\mu \, \xi(r,\mu) \, L_\ell(\mu) ,
\label{eq:multipole}
\end{equation}
where $\xi^\ell$ is the $\ell$th multipole of $\xi$ and $L_\ell$ is the Legendre polynomial of mode $\ell$. The galaxy correlation function multipole components we examine in our analysis are $\xi^0_{gg}(r)$ and $\xi^2_{gg}(r)$.

We also measured the galaxy-velocity cross-correlation function, focusing on the leading-order contribution from the dipole, $\xi^1_{gv}(r)$.  The estimator has the form,
\begin{equation}
    \begin{split}
    \xi_{gv}(r_i,\mu_j) = \left\langle \frac{1}{\alpha} \right\rangle & \left[ \left( \frac{N^R_g N^R_v}{N^D_g N^D_v} \right) D_gD_v - \frac{N^R_g}{N^D_g} D_gR_v \right. \\ & \left. - \frac{N^R_v}{N^D_v} R_gD_v + R_gR_\eta \right] ,
    \end{split}
    \label{eq_xigv}
\end{equation}
where the pair count is defined using the $\eta$ variable introduced in Eq.\ref{eq_eta} rather than velocity, such that,
\begin{equation}
    A_gB_\eta = \frac{\sum_{a,b}^{A,B\in(r_i, \mu_j)} w^g_a \, w^v_b \cos\theta_b \, \eta_b}{\sum_{a,b}^{R,R\in(r_i, \mu_j)} w^g_a \, w^v_b \cos^2\theta_b} ,
\end{equation}
where $\theta_b$ is the angle between the position vector of galaxy $b$ and the separation vector between the galaxy pair $a,b$ being considered, as depicted by Fig.1 in \cite{2023MNRAS.518.2436T}.  The normalisation at the front of Eq.\ref{eq_xigv}, which compensates for the conversion of velocity to $\eta$, is given by,
\begin{equation}
    \left\langle \frac{1}{\alpha} \right\rangle = \frac{\sum_{a,b}^{A,B\in(r_i,\mu_j)} w_a^v \, w_b^v}{\sum_{a,b}^{A,B\in(r_i,\mu_j)} w_a^v \, w_b^v \, \alpha_b} ,
\end{equation}
which transforms the galaxy-$\eta$ correlation to a galaxy-velocity correlation.  The optimal weights used for the velocity sample, when applied to $\eta$ variables, are given by \citep{2023MNRAS.518.2436T},
\begin{equation}
    w^v_a = \frac{1}{\alpha_a \, n^v_a \, P_v + \sigma^2_\eta/\alpha_a} ,
\end{equation}
where $P_v = 10^9 \, h^{-3} \, \textrm{Mpc}^3\textrm{km}^2\textrm{s}^{-2}$ is the characteristic velocity power spectrum amplitude, $n^v_a$ is the local number density of the velocity sample at galaxy $a$, $\sigma_\eta$ is the measurement error for the log-distance variable, and $\alpha$ is defined by Eq.\ref{eq_eta}.  Having measured the galaxy-velocity correlation, we extract the dipole $\xi^1_{gv}(r)$ using Eq.\ref{eq:multipole}.

The radial velocity auto-correlation function can be probed by the $\psi_1$ and $\psi_2$ estimators defined by \cite{1989ApJ...344....1G}, which we again measure in terms of the $\eta$ variables using the estimators,
\begin{equation}
    \psi_x(s) = \left\langle \frac{1}{\alpha^2} \right\rangle \left[ \left( \frac{N^R_v}{N^D_v} \right)^2 DD_{\psi_x} - 2\frac{N^R_v}{N^D_v}DR_{\psi_x} + RR_{\psi_x} \right] ,
\label{eq:psiest}
\end{equation}
where $x = \{ 1, 2 \}$, and in this case we only consider the monopole correlation functions, so we do not need to divide the pair count measurement into bins of $\mu$.  The pair counts are defined as,
\begin{equation}
    AB_{\psi_1}(s) = \frac{\sum_{a,b}^{A,B \in r_i} w^v_a \, w^v_b \cos\theta_{ab} \, \eta_a \, \eta_b}{\sum_{a,b}^{R,R \in r_i} w^v_a \, w^v_b \, \cos^2\theta_{ab}} ,
\end{equation}
\begin{equation}
    AB_{\psi_2}(s) = \frac{\sum_{a,b}^{A,B \in r_i} w^v_a \, w^v_b \, \cos\theta_a \cos\theta_b \, \eta_a \, \eta_b}{\sum_{a,b}^{R,R \in r_i} w^v_a \, w^v_b \, \cos\theta_a \cos\theta_b \cos\theta_{ab}} ,
\end{equation}
where $\theta_{ab}$ is the angle between the position vectors of galaxies $a$ and $b$.  The normalisation term is now,
\begin{equation}
    \left\langle \frac{1}{\alpha^2} \right\rangle = \frac{\sum_{a,b}^{A,B \in r_i} w_a^v \, w_b^v}{\sum_{a,b}^{A,B \in r_i} w_a^v \, w_b^v \, \alpha_a \, \alpha_b} .
\end{equation}

\begin{figure*}
    \centering
    \includegraphics[width=1.8\columnwidth]{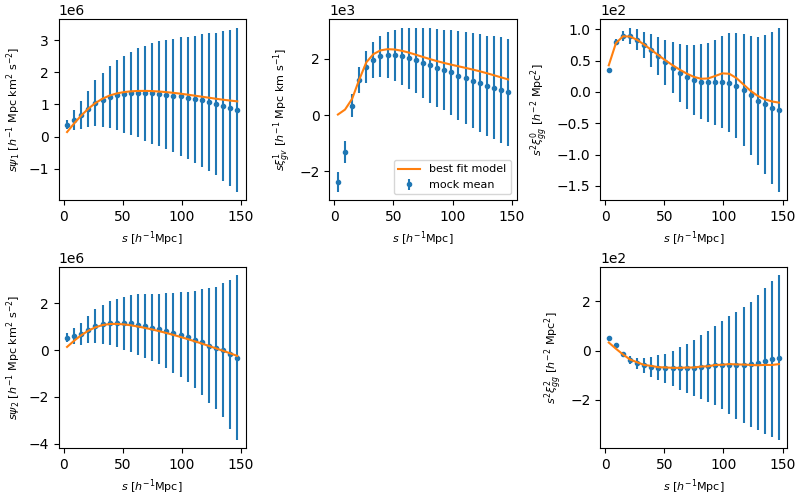}
    \caption{The mean of the five velocity and galaxy correlation functions across the 2048 mocks generated for the SDSS survey. They have been multiplied by factors of the separation scale $s$ to reduce the dynamic range for the convenience of visualisation.} The errors are displayed as the square root of the diagonal of the covariance matrix.  The solid line plots the average of the best fitting theoretical models to each mock.
    \label{fig:corrsdssmock}
\end{figure*}

\begin{figure*}
    \centering
    \includegraphics[width=1.8\columnwidth]{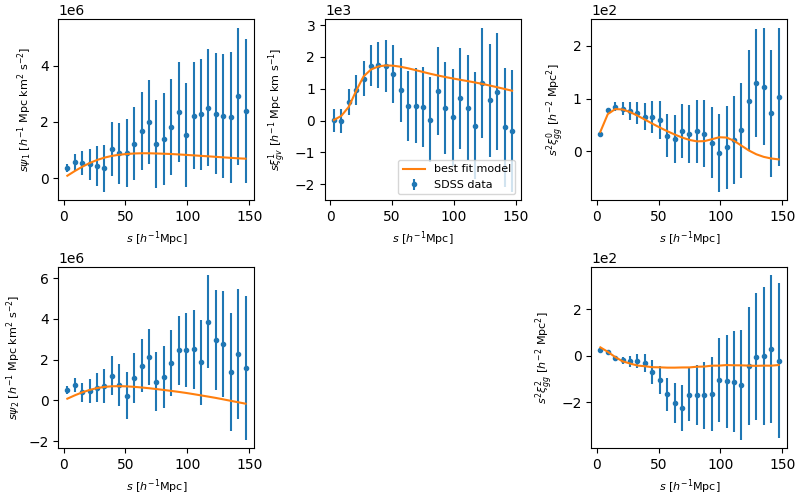}
    \caption{The five velocity and galaxy correlation functions for the SDSS data, multiplied by factors of $s$ to highlight detail}. The errors are displayed as the square root of the diagonal of the covariance matrix, and the best-fitting model is shown as the solid line. Measurements in different separation bins, especially for the velocity correlations, are strongly correlated with each other. As such, the model can be systematically shifted from the data without strongly affecting the $\chi^2$ value.
    \label{fig:corrsdssdata}
\end{figure*}

Our measurements of the five correlation functions used in this study $\{ \psi_1$, $\psi_2$, $\xi_{gv}^1$, $\xi_{gg}^0$, $\xi_{gg}^2 \}$ are displayed for the SDSS datasets in Fig.\ref{fig:corrsdssmock} and Fig.\ref{fig:corrsdssdata}, where we scale the correlation functions by powers of $s$ to reduce their dynamic range for convenience of visualisation.  Fig.\ref{fig:corrsdssmock} shows the average correlation function when these estimators are applied to the SDSS mock catalogues, displaying the mean and standard deviation of the estimators when applied to the simulations. The best-fitting theoretical model with the lowest $\chi^2$ value from Eq.\ref{eq_chi2} is also plotted, showing agreement between the theory and simulation. Fig.\ref{fig:corrsdssdata} displays the correlation functions estimated from the SDSS data, along with the best-fitting theoretical model.  The correlation function measurements for the 6dFGS dataset were already plotted as Fig.6 in \cite{2023MNRAS.518.2436T}, so we do not reproduce them here. 

\subsection{Covariance}
\label{sec:cov}

\begin{figure}
    \centering
    \includegraphics[width=\columnwidth]{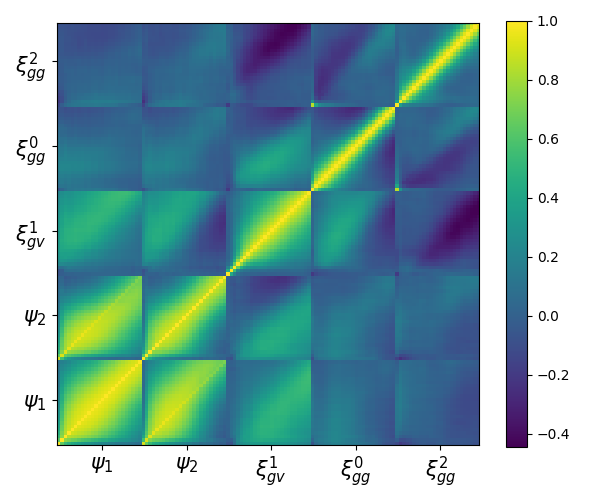}
    \caption{The renormalised covariance matrix calculated for the 2048 SDSS mocks using Eq.\ref{eq_corrmat}. The correlation coefficients are plotted as $C_{ij}/\sqrt{C_{ii} C_{jj}}$.  The data vector is arranged in the order: ($\psi_1$, $\psi_2$, $\xi_{gv}^1$, $\xi_{gg}^0$, $\xi_{gg}^2$).}
    \label{fig:covariance}
\end{figure}

The covariance matrix of the ensemble of correlation functions, across different separations and statistics, was calculated for each survey from the mock measurements as:
\begin{equation}
    C_{ij}=\frac{1}{N-1}\sum_k\left(\xi_{ki}-\overline{\xi}_{i}\right)\left(\xi_{kj}-\overline{\xi}_{j}\right) ,
\label{eq_corrmat}
\end{equation}
where $N$ is the number of mocks, $\xi_{ki}$ are correlation functions for each of the mocks where $k$ iterates through all individual mocks and $i$ refers to each separation bin of each correlation function in the order $\{ \psi_1$, $\psi_2$, $\xi_{gv}^1$, $\xi_{gg}^0$, $\xi_{gg}^2 \}$, and $\overline{\xi}_{i}$ is the mean of the correlation functions across all mocks.  We note that the error in the covariance matrix implied by the finite number of mocks, quantified by the Hartlap factor \citep{2007A&A...464..399H}, is insignificant for our analyses. For the SDSS mock datasets, the resulting covariance matrix is shown in Fig.\ref{fig:covariance}. We highlight the strong correlations between velocity statistics, driven by the large-scale modes \citep{2024MNRAS.527..501B}.  \cite{2023MNRAS.518.2436T} displayed the corresponding covariance matrix for the 6dFGS data in their Fig.2. We use these covariance matrices for fitting the model parameters to the data, as described in Sec.\ref{sec:gr}.

\section{General Relativity fits}
\label{sec:gr}

In this section we describe our model parameter fits to the measured correlation functions assuming General Relativity.  We will focus on fitting the growth rate to the SDSS galaxy and velocity correlation functions, for both the mocks and data.  We note that a similar analysis of the 6dFGS data has already been presented by \cite{2023MNRAS.518.2436T}, so we do not repeat that analysis here, but we will compare the results below.

In the GR fit we vary 3 parameters: the growth rate $f$, the galaxy bias $b$, and the non-linear velocity dispersion $\sigma_v$. We perform these fits at fixed $\sigma_8=0.815$, quoting our results as the combinations $f \sigma_8$ and $b \sigma_8$ which are degenerate in linear theory. From these parameters, the theoretical values of the correlation functions can be calculated as described in Sec.\ref{sec:corrmod}. These can then be compared against the correlation functions measured using the estimators in Sec.\ref{sec:corrmeas}.

We parameterised the goodness of the model fit using the $\chi^2$ statistic,
\begin{equation}
    \chi^2=\sum_{i,j}C_{ij}^{-1} \left( \xi_i^d - \xi_i^t \right) \left( \xi_j^d - \xi_j^t \right) ,
\label{eq_chi2}
\end{equation}
where $\xi^d$ is the correlation function extracted from data and $\xi^t$ is the correlation function predicted from linear theory. $C_{ij}$ is the covariance matrix determined in Sec.\ref{sec:cov}. We summed over separation bins $i$ and $j$ for all correlation functions within the fitting range $s > 20 \, h^{-1}$ Mpc, for which linear theory may be applicable. This choice of fitting range is motivated by the analysis of \cite{2023MNRAS.518.5929L} and \cite{2023MNRAS.518.2436T}, and we tested the dependence of our results on the minimum fitted scale in Sec.\ref{sec:grdata} below.

For a given data set, the $\chi^2$ statistic was determined across a grid of equally-spaced parameter values: $f$ was calculated for 100 values from 0 to 1, $b$ was calculated for 100 values from 0 to 2, and $\sigma_v$ was calculated for 29 values from 20 to 600 km s$^{-1}$. For a GR model we can set $g_0 = 1$.  Assuming the data has Gaussian errors, the likelihood of each model is found from the $\chi^2$ value as,
\begin{equation}
    P \propto e^{-\frac{1}{2}\chi^2} .
\label{eq_prob}
\end{equation}
Assuming a uniform prior for all free parameters across their investigated ranges, the full posterior probability distribution only requires that Eq.\ref{eq_prob} be normalised to sum to $1.0$.

\subsection{GR fit to SDSS mocks}

We first apply this fitting method to the 2048 SDSS mock catalogues. This allows us to test that our model is sufficient to recover the fiducial growth rate in an unbiased fashion, given the statistical errors. For the effective redshift $z_{\rm eff} = 0.073$ and fiducial cosmological model used to generate the mock data, we calculate the theoretical $f\sigma_8$ value as $0.438$.

\begin{figure}
    \centering
    \includegraphics[width=\columnwidth]{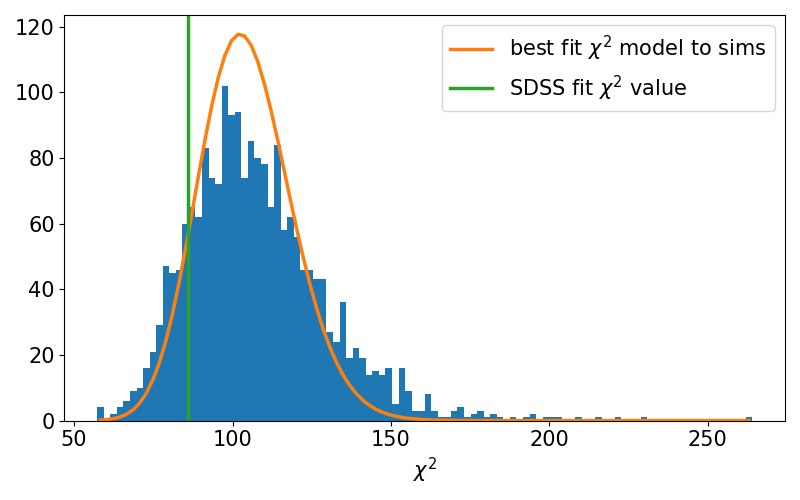}
    \caption{Histogram of the $\chi^2$ values of the best-fitting models for the 2048 SDSS mocks. The best-fitting $\chi^2$ distribution for all best fitting simulation models is overlaid, with a mean value of 104.3. The vertical line indicates the $\chi^2$ value of the fit to the real SDSS data, at a value of 86.0.}
    \label{fig_f4Smocchi}
\end{figure}

A histogram of the $\chi^2$ values of the best-fitting models for each of the 2048 mock surveys can be seen in Fig.\ref{fig_f4Smocchi}. The distribution is best fit by a theoretical $\chi^2$ distribution with a mean value of 104.3, while the number of data points is 110 fit with 3 effective free parameters. Hence, the mean $\chi^2$ closely follows the theoretical prediction, showing that the linear theory used to derive the theoretical correlation function is sufficiently accurate for modeling our measurements and the following statistical analysis.

\begin{figure}
    \centering
    \includegraphics[width=\columnwidth]{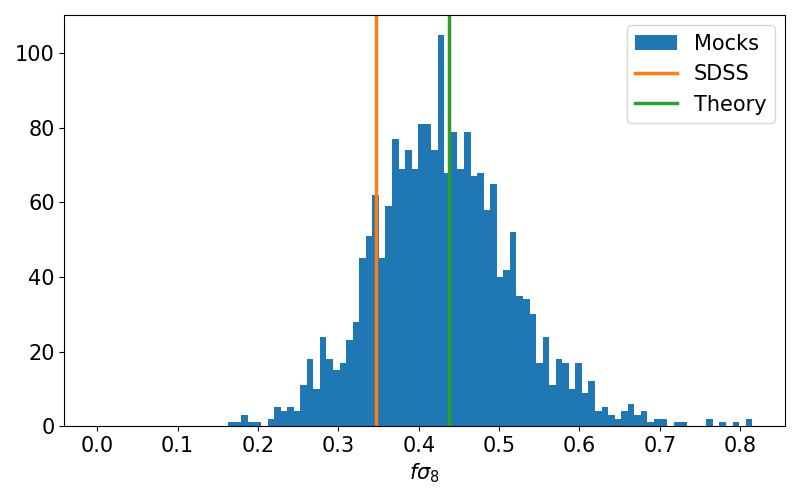}
    \caption{Histogram of the normalised growth rate of structure ($f\sigma_8$) for the best fitting models to all 2048 SDSS mocks. The orange and green vertical lines show the value of the best-fitting model to the real SDSS data, and the value predicted from theory, respectively.}
    \label{fig_f5Smocf}
\end{figure}

Fig.\ref{fig_f5Smocf} shows the histogram of $f\sigma_8$ values extracted from the best-fitting models for each of the SDSS mocks. The mean $f\sigma_8$ value fitted to the mocks is $0.431$, with a standard deviation of $0.087$.  Hence, the mock analysis successfully recovers the fiducial growth rate value.

\subsection{GR fit to SDSS data}
\label{sec:grdata}

\begin{figure}
    \centering
    \includegraphics[width=\columnwidth]{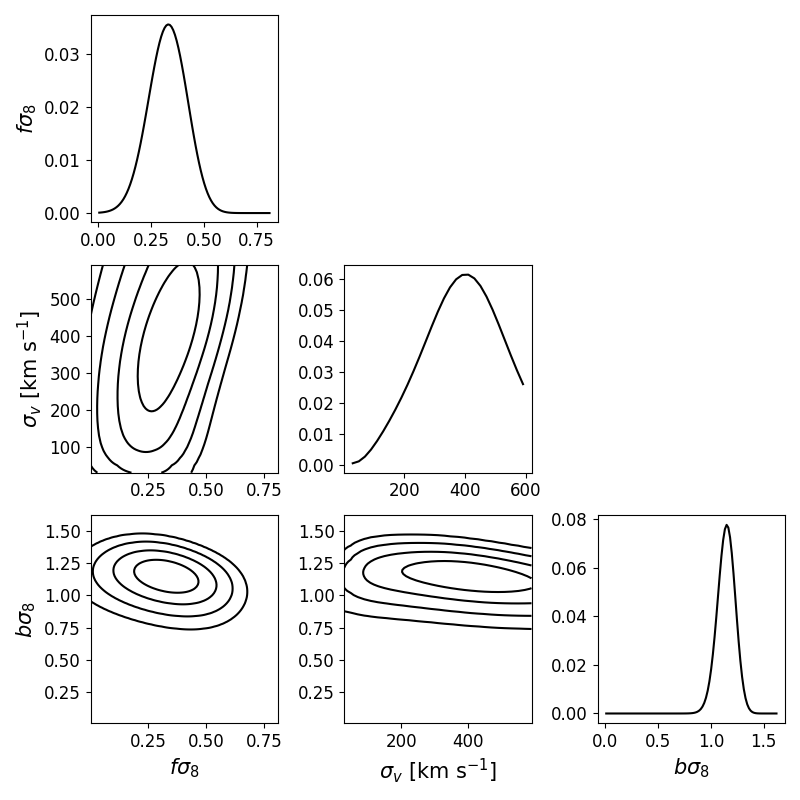}
    \caption{The joint confidence regions of the three fitted model parameters $(f\sigma_8, b\sigma_8, \sigma_v)$ for the fit to the SDSS data set.  The contours indicate 1 to 4 sigma confidence regions, and the plot also shows the posterior probability distribution of each individual parameter.}
    \label{fig_f7Sdat}
\end{figure}

We then applied our method to fit the correlation function dataset from SDSS. The joint confidence regions of all the model parameters are displayed in Fig.\ref{fig_f7Sdat}. The model that provides the best fit to the data has $f\sigma_8 = 0.346$, $b\sigma_8 = 1.15$ and $\sigma_v = 410$ km s$^{-1}$. With 110 data points, the fit has $\chi^2=86.0$. The posterior probability distribution has a median value and $68\%$ confidence regions of $f\sigma_8 = 0.329^{+0.081}_{-0.083}$, $b\sigma_8 = 1.15^{+0.074}_{-0.078}$ and $\sigma_v = 385^{+111}_{-126}$ km s$^{-1}$.  The General Relativity prediction is narrowly outside the 1-$\sigma$ confidence interval of our fit.

\begin{figure}
    \centering
    \includegraphics[width=\columnwidth]{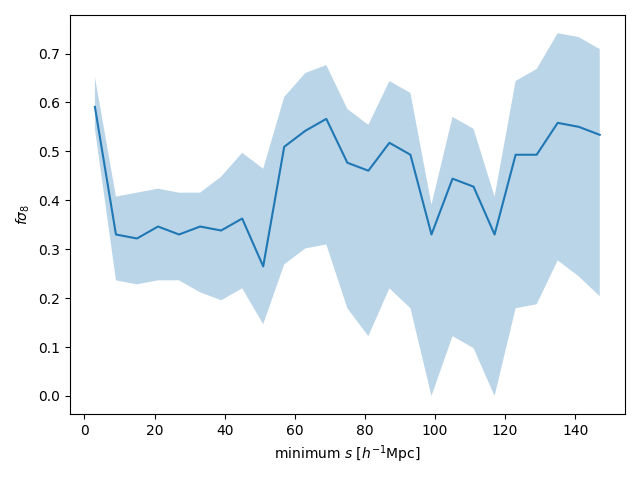}
    \caption{The dependence of the best-fitting growth rate of structure value on the minimum scale used in fitting. The solid line shows the most likely growth rate value, and the shaded region shows the 1-$\sigma$ confidence region from the posterior probability distribution.}
    \label{fig_minscale}
\end{figure}

We assumed a minimum fitting scale of 20 $h^{-1}$Mpc in our analysis, in order to exclude the effects of non-linearities.  To ensure the validity of this choice, in Fig.\ref{fig_minscale} we show the result of repeating our analysis, varying the minimum fitted scale across our separation range.  Very similar results are obtained for minimum fitted scales in the range  10-40 $h^{-1}$Mpc.  The measured growth rate shows evidence of systematic bias when scales less than 10 $h^{-1}$Mpc are included in the fit, and the error in the measurements significantly increases for larger minimum fitted scales.  We also performed similar tests that our conclusions when constraining modified gravity parameters in later sections did not significantly depend on the fitting range.

Our measurement agrees with previous related measurements of the local normalised growth rate within the margin of error, which we summarise here.  \cite{2023MNRAS.518.1840L} fitted a similar model to the SDSS PV sample using a maximum-likelihood method applied to the full density and velocity fields, finding $f\sigma_8 = 0.405_{-0.071}^{+0.076}$, within the statistical margin of error of our measurement. Now considering other peculiar velocity surveys: \cite{2023MNRAS.518.2436T} conducted the same correlation-function fitting process on the 6dFGS dataset, finding $f \sigma_8 = 0.358 \pm 0.075$.  \cite{2014MNRAS.444.3926J} fit the correlation of the peculiar velocity field of the 6dFGS and a sample of supernova, finding a value $f\sigma_8=0.418\pm0.065$. \cite{2017MNRAS.471.3135H} used the peculiar velocity power spectrum to fit the 2MASS Tully-Fisher survey (2MTF) and found $f\sigma_8=0.51^{+0.09}_{-0.17}$ and \cite{2019MNRAS.487.5235Q} fit the density and momentum power spectra of both the 6dFGS and 2MTF datasets, recovering a value of $f\sigma_8=0.404^{+0.082}_{-0.081}$. Finally, \cite{2020MNRAS.494.3275A} performed a joint maximum-likelihood study of the overdensity and velocity fields of 6dFGS, determining $f\sigma_8 = 0.384 \pm 0.052$.  We summarise these findings by noting that a series of different methodologies produce similar determinations of the local growth rate of structure from peculiar velocity measurements, which show no significant evidence for departures from the General Relativity prediction.  In the next section, we consider what limits can be placed on modified gravity theories by these same datasets.

\section{Modified Gravity fits}
\label{sec:modgrav}

We now compare both the SDSS and 6dFGS datasets to theoretical correlation functions produced for DGP and $f(R)$ modified gravity models.

\subsection{Modified Gravity considerations}
\label{sec:mg}

In modified gravity fits, the $g_0$ value introduced in Sec.\ref{sec:modgrowth}, which normalises the growth factor, can no longer be assumed to be equal to $1.0$ without further information, due to the growth rate history potentially diverging from the $\Lambda$CDM model. Since the growth factor enters the matter power spectrum as a quadratic factor, all theoretical correlation functions would be multiplied by $g_0^2$ as this normalisation varies.

The normalisation of the matter power spectrum can be well-constrained by CMB measurements in a manner independent of the late-time Universe, which allows us to introduce a prior in $g_0$.  We used the prior determined by \cite{2023PhRvD.107j3505L}, who determine a primordial measurement of the combination of the power spectrum amplitude $A_s$ and reionisation optical depth $\tau$, $A_s e^{-2\tau} = (1.873 \pm 0.012) \times 10^{-9}$ from {\it Planck} CMB measurements. For our choice of fiducial cosmological model, the reionisation optical depth is $\tau=0.0829$ and the spectrum amplitude is $A_s = 2.23\times10^{-9}$.  Therefore, since $A_s$ is proportional to $g_0^2$, this constraint corresponds to a Gaussian prior in $g_0^2$ with a mean of $0.9874$ and standard deviation of $0.00633$.

Since the correlation functions simply depend on $g_0^2$, we can analytically marginalise over $g_0$ without calculating the more computationally-intensive probability over a grid. This is performed by integrating the parameter $g_0^2$ across Eq.\ref{eq_chi2}, multiplied by its Gaussian prior. The result of this process, as described in \cite{2002MNRAS.335.1193B}, can be characterised as an alteration within the $\chi^2$ computation of Eq.\ref{eq_chi2} to use the inverse covariance matrix given as Eq.\ref{eq_invcoralt} below, evaluating the theoretical correlations at the mean of the $g_0^2$ prior:
\begin{equation}
    (C_g^{-1})_{ij} = C^{-1}_{ij} -\frac{\sum_{k,l} \left( C^{-1}_{ik} \, \xi^t_k \, \xi^t_l \, C^{-1}_{lj} \right)}{\sum_{m,n} \left( \xi^t_m \, C^{-1}_{mn} \, \xi^t_n \, \right) + \sigma_g^{-2}} ,
    \label{eq_invcoralt}
\end{equation}
where $C^{-1}$ is the original inverse of the covariance matrix, $C^{-1}_g$ is the altered inverse covariance matrix after analytical marginalisation that will be used in Eq.\ref{eq_chi2} for fitting modified gravity models, $\sigma_g$ is the standard deviation of the Gaussian prior in $g_0^2$, and $\xi^t$ is the theoretical correlation function evaluated at the mean of the prior.

The characteristic parameters of the modified gravity models we consider in this study, $f_{R0}$ and $r_c$, asymptotically converge towards $\Lambda$CDM models as $f_{R0} \rightarrow 0$ and $r_c \rightarrow \infty$.  The limits we quote for deviations from $\Lambda$CDM depend on the parameterisations and priors we adopt for these variables. For DGP models our fiducial choice is a uniform prior in $r_c$ from $0.01$ to $100$ in units of $c/H_0$. We also consider parameterising these models by $\Omega_{r_c}$, an effective normalised cosmic energy density defined by,
\begin{equation}
\Omega_{r_c} = \frac{c^2}{4H_0^2r_c^2} ,
\end{equation}
where $\Omega_{r_c}=0$ retrieves $\Lambda$CDM \citep{2007ApJ...666..716D}.  In this case, we use a flat prior in $\Omega_{r_c}$ over the equivalent range from $2.5\times10^{-5}$ to $2500$. Other prior options are presented in Table \ref{tab:MGlimits} and discussed further below.  For $f(R)$ models, we compare two priors.  Our fiducial choice is a uniform prior in $-\log_{10}{|f_{R0}|}$ between $-1.0$ and $10.0$.  We compare this with a uniform prior in $f_{R0}$ over the same range.  In all cases, we quote 2$\sigma$ confidence level limits for deviations of these parameters from the $\Lambda$CDM limit, for each prior choice.

\subsection{Fits to mock surveys}

\begin{figure}
    \centering
    \includegraphics[width=\columnwidth]{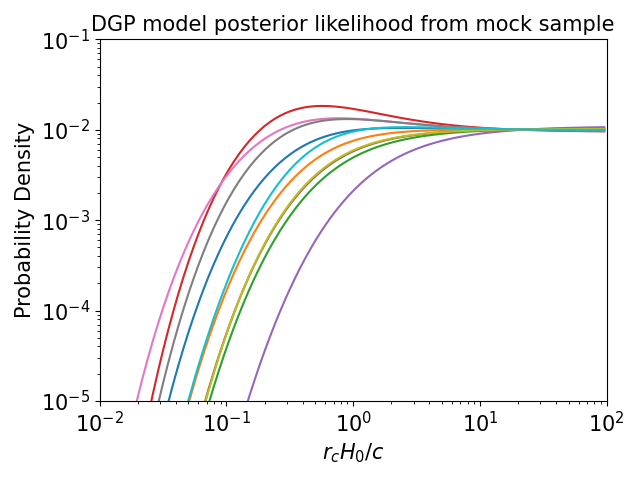}
    \includegraphics[width=\columnwidth]{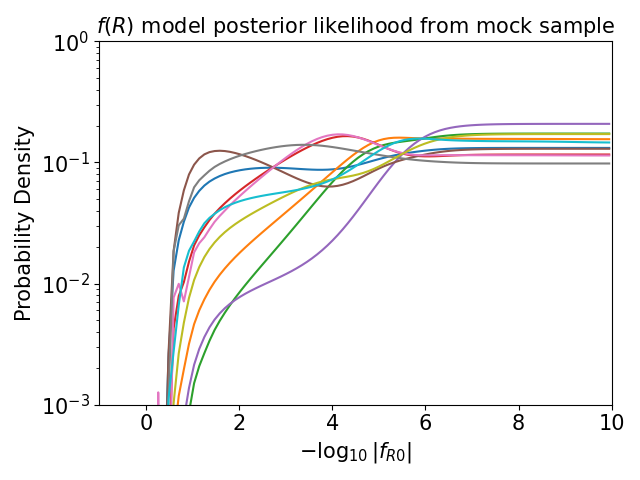}
    \caption{The posterior probability distributions for the variable used to characterise the DGP and $f(R)$ modified gravity models, for 10 randomly-chosen $\Lambda$CDM mock data sets generated for the SDSS survey. The other model parameters ($g_0$, $b$ and $\sigma_v$) are marginalised.  We assume uniform priors in $r_c$ and $\log_{10}{|f_{R0}|}$ in this analysis. As both parameters diverge to infinity, the model converges to $\Lambda$CDM. It can be seen that the posteriors generally show an asymptotic likelihood as the fitted parameters approach $\Lambda$CDM, which is favoured in the fits.}
    \label{fig:p2f8Smocks}
\end{figure}

We start by fitting both DGP and $f(R)$ models to the mock correlation function measurements of the 6dFGS and SDSS datasets.  The posterior probabilities found from fitting DGP and $f(R)$ models to a random subset of mock data samples, for our fiducial choice of uniform priors in $r_c$ and $\log_{10}{|f_{R0}|}$, respectively, can be seen in Fig.\ref{fig:p2f8Smocks}.  These results help to validate this analysis technique, as the simulations have known initial conditions and gravity laws constructed from the $\Lambda$CDM model.  As can be seen in Fig.\ref{fig:p2f8Smocks}, the mock measurements return fitting results that conform with this model, with the most likely models found at the  parameter values closest to $\Lambda$CDM (where the posteriors converge, since the models become asymptotically close to $\Lambda$CDM).

\begin{figure}
    \centering
    \includegraphics[width=\columnwidth]{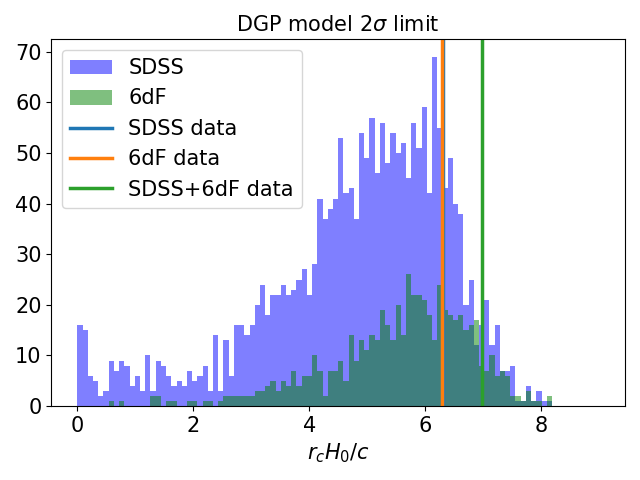}
    \includegraphics[width=\columnwidth]{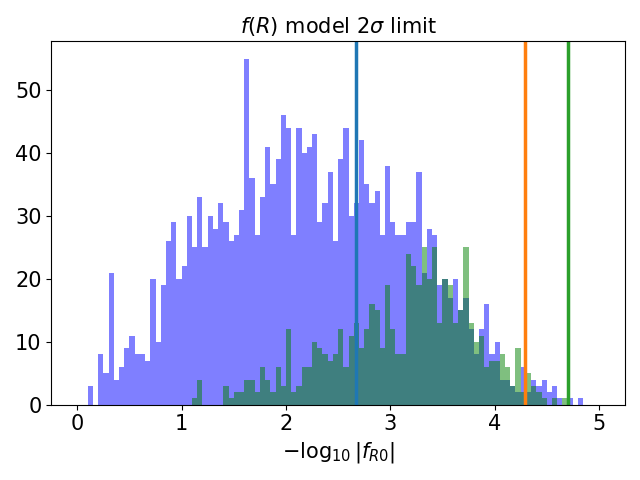}
    \caption{Histogram of the 2-$\sigma$ limits on the DGP and $f(R)$ models for the 2048 SDSS mocks and the 600 6dFGS mocks. For the DGP constraint a uniform prior in $r_c$ from $0.01$ to $100$ in units of $c/H_0$ was used, and the $f(R)$ constraint used a uniform prior in $\log{|f_{R0}|}$ from $-1$ to $10$.  The 2-$\sigma$ constraints found from the real data surveys are plotted as vertical lines (for each individual survey, and the combination).}
    \label{fig:p2f10rcmochist}
\end{figure}

The 2-$\sigma$ confidence limits for the modified gravity parameters, enclosing $95\%$ probability toward the $\Lambda$CDM limit, are shown as a histogram for the ensemble of mocks in Fig.\ref{fig:p2f10rcmochist} (again, assuming uniform priors in $r_c$ and $\log_{10}{|f_{R0}|}$). This shows what limits can be placed on deviations from a $\Lambda$CDM universe, given datasets statistically sampled from this model.

\subsection{Fits to data}

\begin{table}
    \centering
    \begin{tabular}{c|c|c|c|c}
    \hline
    Parameter & Prior & SDSS & 6dFGS & Comb.\\
    \hline
    $r_c H_0/c$ & Uniform $\Omega_{r_c}$ & 0.0855 & 0.0957 & 0.209 \\
    $r_c H_0/c$ & Uniform $r_c$ & 6.314 & 6.291 & 6.987 \\
    $r_c H_0/c$ & Uniform $\ln(r_c)$ & 0.657 & 0.732 & 1.076\\
    $r_c H_0/c$ & Uniform $r_c^{-1}$ & 0.175 & 0.200 & 0.367\\
    \hline
    $-\log_{10}|f_{R0}|$ & Uniform $f_{R0}$ & 0.664 & 0.794 & 0.994 \\
    $-\log_{10}|f_{R0}|$ & Uniform $\log_{10}|f_{R0}|$ & 2.669 & 4.291 & 4.703 \\
    \hline
    \end{tabular}
    \caption{The 2-$\sigma$ lower limits on the modified gravity parameters, for different choices of the prior.  These limits were found using the SDSS, 6dFGS and combined datasets.  The upper half of the table displays the bounds for $r_c H_0/c$ for uniform priors in $\Omega_{r_c}$, $r_c$, $\ln(r_c)$, and $r_c^{-1}$ (to demonstrate their dependence on this consideration). The lower half of the table displays the bounds for $-\log_{10}|f_{R0}|$, for uniform priors in $f_{R0}$ and $\log_{10}|f_{R0}|$.}
    \label{tab:MGlimits}
\end{table}

\begin{figure}
    \centering
    \includegraphics[width=\columnwidth]{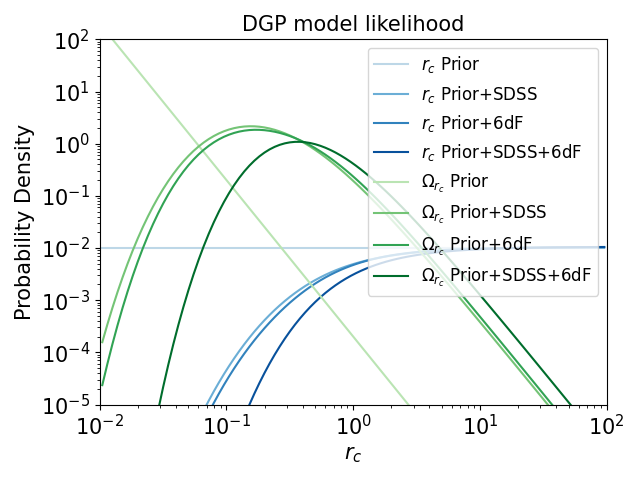}
    \includegraphics[width=\columnwidth]{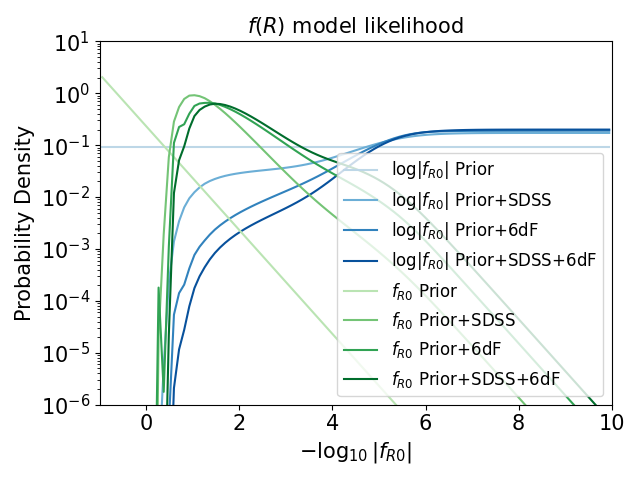}
    \caption{The posterior probability distributions derived from the 6dFGS and SDSS datasets for the modified gravity model parameters $r_c H_0/c$ (top) and $-\log_{10}|f_{R0}|$ (bottom), marginalising over $g_0$, $b$, and $\sigma_v$. The effective prior and the posterior of the combined data set is also shown. A uniform prior in $r_c$ (top) and $\log_{10}|f_{R0}|$ (bottom) are used for the blue posterior curve, while the results found with an alternate uniform prior in $\Omega_{r_c}$ (top) and $f_{R0}$ (bottom) are shown in green.}
    \label{fig:p2f11DGPdata}
\end{figure}

We now apply our fitting method to the correlation functions measured from the 6dFGS and SDSS datasets.  The posterior probability distributions resulting from the fitting process can be seen in Fig.\ref{fig:p2f11DGPdata} for various different choices of prior, clarifying the additional constraining information with respect to this prior provided by the likelihood of the data. The best-fitting DGP model has $r_cH_0/c=95.60$ with $\chi^2=86.4$, and the best $f(R)$ model has $-\log_{10}|f_{R0}|=9.395$ with $\chi^2=87.1$. Both best-fitting modified models are very close to the most GR convergent model in the range of parameters we considered and neither showed better statistical fitting than the GR model at a $\chi^2$ value of $86.0$.  Our fits hence do not show a statistical preference for the modified gravity scenario over GR.

We also calculated 2-$\sigma$ lower limits for each case, which are recorded in Table \ref{tab:MGlimits}.  For DGP models, the lower limits in $r_c H_0/c$ are $6.291$ (6dFGS) and $6.314$ (SDSS), assuming a uniform prior in $r_c$.  For $f(R)$ models, the lower limits in $-\log_{10}(|f_{R0}|)$ are $4.291$ (6dFGS) and $2.669$ (SDSS), assuming a uniform prior in $\log_{10}|f_{R0}|$.  As the 6dFGS and SDSS surveys cover separate regions of the sky, we can make the assumption that they are independent datasets, such that the relative probability of any given model is given by the product of both posterior likelihoods. In this case, the combined 6dFGS and SDSS datasets constrain DGP and $f(R)$ models with limits of $r_c H_0/c > 6.987$ and $-\log_{10}(|f_{R0}|) > 4.703$.

In Fig.\ref{fig:p2f11DGPdata} we also present the posterior probability distributions inferred for the modified gravity parameters assuming the alternative priors discussed in Sec.\ref{sec:mg}, demonstrating that this choice has a substantial effect on our derived limits.  The $2\sigma$ limits found for all these cases are recorded in Table \ref{tab:MGlimits}.  However, in no case do we find a preference for a modified gravity scenario compared to the GR model.

We now compare our constraints to existing limits in the literature.  Using the galaxy correlation function multipoles from SDSS Data Release 7, \cite{2013MNRAS.436...89R} obtained a $2\sigma$ bound of $r_cH_0/c > 0.076$ assuming a uniform prior in $r_c^{-1}$. \cite{2016PhRvD..94h4022B} also used these galaxy clustering correlations to fit to SDSS Data Release 12 with a uniform prior in $r_c^{-1}$ to obtain a $2\sigma$ bound of $r_cH_0/c > 1.03$.  Whilst comparison of different limits is complicated by the choice of prior, using an equivalent prior with our data results in a limit intermediate between these cases (see Table \ref{tab:MGlimits}).

For $f(R)$ gravity,  \cite{2012PhRvD..85l4038L} determined a $2\sigma$ bound $-\log_{10}|f_{R0}| > 3.72$ from a suite of data including supernovae, baryon acoustic oscillations, the Hubble constant and Cosmic Microwave Background data. \cite{2015PhRvD..92d4009C} used galaxy cluster surveys together with similar cosmological datasets as above to place the $2\sigma$ bound $-\log_{10}|f_{R0}| > 4.79$. Finally, \cite{2016PhRvL.117e1101L} determined the $2\sigma$ bound $-\log_{10}|f_{R0}| > 5.16$ using weak lensing surveys and the CMB.  Whilst we again caution that the choice of prior is significant in setting these limits, we find that peculiar velocity surveys can place competitive constraints.

\section{Conclusion}
\label{sec:conc}

The nature of dark energy and explanation of the expansion history of the universe remains one of the biggest puzzles in astrophysics.  Together with the cosmic expansion, the physical nature of dark energy influences the growth and assembly of matter under gravity on smaller scales.  To gain further insight into this question, we have investigated the growth of cosmic structure using correlations between direct observations of galaxy peculiar velocities, and the surrounding galaxy density field. In the large-scale linear regime, the growth of structure and effective strength of gravity are directly linked to the peculiar velocity distribution.  Large surveys of peculiar velocities can therefore offer information on the growth rate of structure across a variety of cosmic scales, through the use of two-point correlation functions.

In this paper we have used the parallel and perpendicular components of the velocity auto-correlation, the galaxy-velocity cross-correlation, and the monopole and quadrupole of the galaxy auto-correlation functions in redshift space as probes to capture information about the growth rate of structure.  We analytically calculated these correlations for a range of different models, focusing on the DGP and Hu-Sawicki $f(R)$ gravity models as representative of broad classes of potential theories. We parameterised these models in terms of their galaxy bias, non-linear velocity dispersion, and characteristic parameter defining their deviation from General Relativity. For DGP the characteristic parameter is the crossover length $r_c$, and for $f(R)$ gravity it is the amplitude of the deviating action term characterised by $-\log_{10}|f_{R0}|$.

We measured these correlation functions from the 6dFGS velocity dataset and recent SDSS peculiar velocity survey, the two largest current samples of their kind, as well as a number of simulated mock datasets created to mimic the clustering statistics of the real surveys, which we used to determine the data covariance and perform validation tests. These correlations were fit against a broad range of possible theoretical correlation functions for our gravity models, using maximum likelihood methods.

Our analysis did not detect any significant deviations from the predictions of the $\Lambda$CDM cosmological model, whilst placing significant new limits on such deviations.  Under the assumption of a GR model, we presented a new determination of the growth rate of structure, $f\sigma_8 = 0.329^{+0.081}_{-0.083}$, obtained from measurements of our suite of galaxy and velocity correlation functions, using SDSS data. This result agrees (within the statistical confidence limits) with the $\Lambda$CDM prediction and other measurements of the local growth rate.  We also found 2-$\sigma$ lower limits, based on a combination of both 6dFGS and SDSS data, of $r_c H_0/c > 6.987$ for DGP, and $-\log_{10}(|f_{R0}|) > 4.703$ for $f(R)$ gravity, assuming uniform priors in $r_c$ and $\log_{10}|f_{R0}|$, where we note that inferences depend significantly on the adopted prior (analysis space) for these parameters.  Comparing these limits with others in the literature, we find that galaxy peculiar velocities can place competitive constraints on these scenarios.

In the future, large galaxy surveys conducted by instruments such as the Dark Energy Spectroscopic Instrument \citep{2023MNRAS.525.1106S}, the 4-metre Multi-Object Spectroscopic Telescope (4MOST) Hemisphere Survey \citep{2023Msngr.190...46T}, the Vera Rubin Observatory \citep{2017ApJ...847..128H} and the Australian Square Kilometre Array Pathfinder WALLABY survey \citep{2023MNRAS.519.4589C}, will provide an order of magnitude more peculiar velocity measurements. With such a wealth of information, analyses such as those we have presented in this paper will be important tools in probing and constraining the possible models that describe the large-scale evolution of our universe.

\section*{Acknowledgments}

We thank the anonymous referee for useful comments which helped us clarify the paper.  This project received financial support through an Australian Government Research Training Program Scholarship awarded to SL.  RJT is supported by the Australian Government through Australian Research Council Discovery Project DP220101610.

\section*{Data Availability}

The data underlying this article will be shared on reasonable request to the corresponding author.

\bibliographystyle{mnras}
\bibliography{main}

\bsp
\label{lastpage}
\end{document}